\documentclass[reprint,aps,prl,showpacs,floatfix,notitlepage,superscriptaddress]{revtex4-1}
\usepackage{epsfig,graphics,amssymb,amsmath,subeqnarray,color,bm,bbm}
\usepackage[toc,page]{appendix}
\usepackage{mathtools}
\usepackage{xcolor}
\usepackage{float}
\usepackage{graphicx}
\usepackage[colorlinks=true,allcolors=blue]{hyperref}

\begin{document}
\title{Exact phoretic interaction of two chemically-active particles}
\author{Babak Nasouri}
\affiliation{Max Planck Institute for Dynamics and Self-Organization (MPIDS), 37077 Goettingen, Germany}
\author{Ramin Golestanian}
\email{ramin.golestanian@ds.mpg.de}
\affiliation{Max Planck Institute for Dynamics and Self-Organization (MPIDS), 37077 Goettingen, Germany}
\affiliation{Rudolf Peierls Centre for Theoretical Physics, University of Oxford, Oxford OX1 3PU, United Kingdom}
\date{\today}

\begin{abstract} 
We study the nonequilibrium interaction of two isotropic chemically-active particles taking into account the exact near-field chemical interactions as well as hydrodynamic interactions. We identify regions in the parameter space wherein the dynamical system describing the two particles can have a fixed-point---a phenomenon that cannot be captured under the far-field approximation. We find that due to near-field effects, the particles may reach a stable equilibrium at a nonzero gap size, or make a complex that can dissociate in the presence of sufficiently strong noise. We explicitly show that the near-field effects are originated from a self-generated neighbor-reflected chemical gradient, similar to interactions of a self-propelling phoretic particle and a flat substrate.
\end{abstract}

\maketitle
Nonequilibrium interfacial transport processes known as phoretic mechanisms \cite{anderson1989,julicher2009} have been key to the development of the field of active matter \cite{RG-LesHouches}. Due to their force-free nature, they have been identified as suitable mechanisms for designing self-propelled active colloids \cite{golestanian2005}, which provide prototypes for the so-called active Brownian particle that serves the role of injecting energy at the small scale in active matter systems \cite{Ramaswamy:2010,Bechinger:2016,elgeti2015}. Moreover, the phoretic activity mediates nonequilibrium interactions between different active colloids via the dynamically generated gradients, leading to a wealth of collective phenomena such as cluster formation and phase separation \cite{cecil12,Golestanian:2012,Palacci2013,saha2014,holger14,Colberg2017,stark2018,Liebchen2015,Kanso2019}. From the biological perspective, prokaryotic \cite{keller2006} and eukaryotic cells \cite{friedl2009} are known to undergo chemotaxis by coupling chemical gradient-sensing to motility. However, it has recently emerged that single enzymes can exhibit a similar chemotactic response \cite{jee2017,zhao2017}, due to a coupling between their chemical nonequilibrium activity and phoretic mechanisms \cite{agud18}, highlighting the relevance of these processes at the molecular scale.
 
Studies of phoretic interactions in many-particle systems has led to a number of nontrivial scenarios for self-organization that gives rise to emergent swimming of clusters of isotropic particles \cite{soto2014,soto2015,palberg17,Niu2018,varma2018} or comet-like propulsion of large swarms \cite{cohen2014,canalejo2019}. These self-organized structures typically involve active colloids in close proximity, where near-field effects play a dominant role. From systematic experimental characterization of isolated catalytically active particles \cite{hows07,Ebbens2012} and the flow-fields generated by them \cite{Ebbens2018}, it is known that various mechanistic details, such as ionic conditions \cite{Ebbens2014,Brown2014} and interactions with nearby surfaces \cite{Das2015,Ibrahim2015,Uspal2015,mozaffari2016,Bayati2019}, are important ingredients for understanding the interactions in many-particle systems.

The first crucial step for studying such complex phenomena is to look at the nonequilibrium interaction between two particles. In particular, resolving the near-field effects arising due to chemical and hydrodynamic interactions will be important for understanding the clustering of phoretic particles. It has already been shown that for a system of two spheres (isotropically-coated \cite{reigh2015,michelin2015b} or Janus \cite{popescu2011,michelin2017}) with no relative motion, accounting for the near-field effects can lead to qualitatively different behavior as compared with the far-field predictions.
It will thus be important to study the exact nonequilibrium interaction between two phoretically-active particles, resolving the near-field effects of the chemical activity as well as the hydrodynamic interactions. This is the task we set out to do here.

\begin{figure*}[t]
\begin{center}
\includegraphics[width=\textwidth]{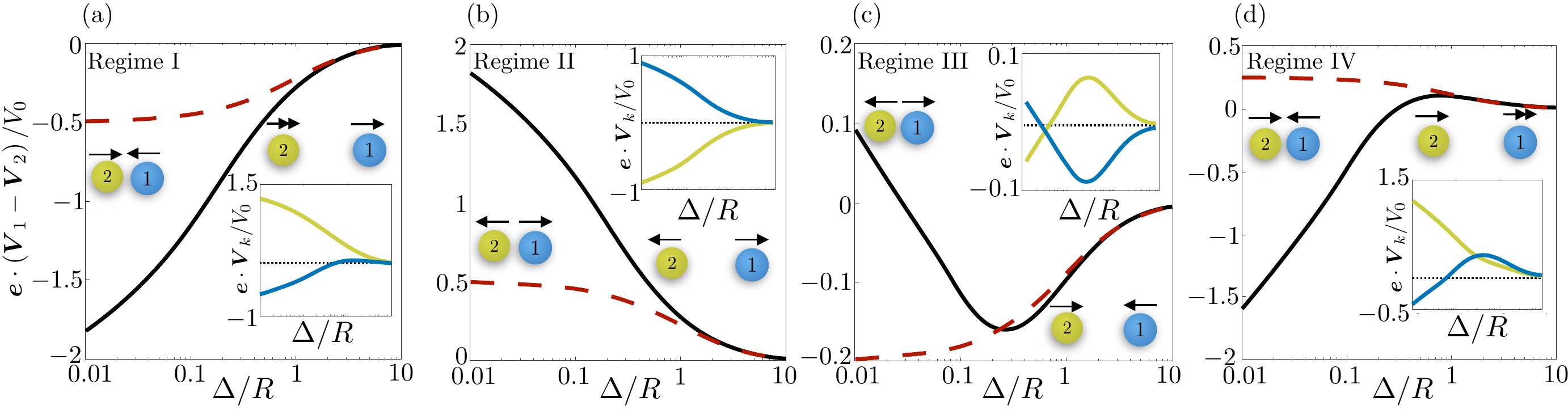}
\caption{Examples for the behavior of the relative velocity versus the gap size in different regimes. The solid lines correspond to the exact solution and the dashed lines represent the far-field approximation. The parameter sets are as follows: (a) $\tilde{\alpha}_1=-3$, $\tilde{\alpha}_2=\tilde{\mu}_1=\tilde{\mu}_2=1$; (b) $\tilde{\alpha}_1=\tilde{\alpha}_2=\tilde{\mu}_1=\tilde{\mu}_2=1$; (c) $\tilde{\alpha}_1=\tilde{\mu}_1=-0.4$, $\tilde{\alpha}_2=\tilde{\mu}_2=1$; (d) $\tilde{\alpha}_1=-2$, $\tilde{\mu}_1=3$, $\tilde{\alpha}_2=\tilde{\mu}_2=1$. Each arrow shows the direction of the velocity for the corresponding particle, and double arrows indicate the larger speed. The velocities of individual particles are shown in the insets; the dotted lines show the value zero. See ancillary files for videos comparing our exact approach and those of far-field approximation in different regimes.}
\label{regimes}
\end{center}
\end{figure*}

Consider such a nonequilibrium system with two particles of radius $R$ interacting with a chemical with diffusion coefficient $D$. Each particle creates a chemical field characterized by its activity $\alpha_k$, which reflects the rate of nonequilibrium catalytic activity, and responds to a chemical gradient via its mobility $\mu_k$, which is controlled by the interfacial interactions \cite{RG-LesHouches,gole07}. By neglecting the hydrodynamic interactions and the near-field chemical interactions (i.e. the `far-field' solution), the velocities of the particles can be found as $\bm{V}_1= \bm{e}\, \alpha_2\mu_1R^2/\left[D (\Delta+2R)^2\right]$ and $\bm{V}_2=- \bm{e}\, \alpha_1\mu_2R^2/\left[D (\Delta+2R)^2\right]$, where $\Delta$ is the clearance between the particles, and $\bm{e}$ is a unit vector pointing from 2 to 1. It is evident that in this system the nonequilibrium activity manifests itself as broken action-reaction symmetry \cite{soto2014,soto2015,canalejo2019}.

For a system of two particles with $\alpha_2\mu_1+\alpha_1\mu_2\neq0$, there can be two scenarios for the relative motion within the far-field description: the interaction is either strictly attractive (leading the particles to form a complex) or strictly repulsive (pushing them apart indefinitely). Here, we show that when exact chemical and hydrodynamic interactions are taken into account, two additional scenarios can exist due to the emergence of a new fixed-point in the effective dynamical system for $\Delta$. The fixed-point can be stable, indicating that the particles will form a complex with a nonzero gap size between them in stationary state, or unstable, meaning that a barrier emerges that needs to be overcome (via thermal activation) for a complex to dissociate. 
For the case of an inert-active pair-interaction, \citet{reigh2018} have shown that a stable fixed-point can emerge, if the active particle has a concentration-dependent surface activity. Here we show that this condition is not necessary, and the interactions of two purely-isotropic particles can remarkably induce a stable fixed-point in the system. This stable state is fundamentally different from the `stationary state' observed in interactions of Janus particles \cite{sharifi2016}, as here the particles do not become stationary; upon reaching the equilibrium gap size, they move together with a mutual nonzero velocity. The existence of an unstable fixed-point in a phoretic pair-interaction has not been discussed before to the best of our knowledge. We also show that these fixed-points can be captured even when the hydrodynamic interactions are neglected, provided the near-field chemical effects are taken into the account.

We formulate the problem for the fluid motion and the solute transport in the regime where inertia and advection can be neglected. The concentration field of the chemical $C$ thereby follows the steady-state diffusion equation $\nabla^2 C=0$, subject to constant flux boundary condition on the surface ${\cal S}_k$ of the $k$th sphere ($k\in\{1,2\}$), represented as $-D \bm{n}_k\cdot\boldsymbol{\nabla}C|_{{\cal S}_k}=\alpha_k$ where $\bm{n}_k$ is a unit vector normal to the surface of sphere $k$. In the absence of any background concentration gradient (or nearby boundaries), we have $C(|\bm{x}|\rightarrow\infty)=0$. The perturbation in the concentration field caused by the particles imposes a local relative slip velocity on the surface of each particle as $\bm{v}^{\rm s}_k=\mu_k\left(\bm{I}-\bm{n}_k\bm{n}_k\right)\cdot\boldsymbol{\nabla}C|_{{\cal S}_k}$. These chemically-induced slip velocities alter the flow field surrounding the particles and may result in their translational motion. To find the instantaneous velocities, one needs to solve the Stokes equations ($\boldsymbol{\nabla} \cdot {\bm \sigma}=\boldsymbol{0}$ and $\boldsymbol{\nabla}\cdot\bm{v}=0$), where $\bm{v}$ is the velocity field that vanishes at infinity and ${\bm \sigma}$ is the stress field. The Stokes equations need to be solved subject to boundary conditions $\bm{v}|_{{\cal S}_k}=\bm{V}_k+\bm{v}^{\rm s}_k$, where $\bm{V}_k$ is the translational velocity of the $k$th sphere. Note that rotation is not considered here due to the axisymmetric nature of the system. In what follows, we take $\alpha_0$ and $\mu_0$ as characteristic values, and define activity and mobility valences as $\tilde{\alpha}_k=\alpha_k/\alpha_0$ and $\tilde{\mu}_k=\mu_k/\mu_0$, as well as a velocity scale $V_0=\alpha_0\mu_0/D$.

The governing equations in the chemical field can be solved exactly using the bispherical coordinate system (see e.g. Refs. \cite{popescu2011,reigh2015,michelin2015b,reigh2018}). Instead of directly solving the Stokes equations for such a system, we employ the Lorentz reciprocal theorem \cite{Lorentz1896,happel1981,stone1996,elfring2017,nasouri2018,masoud2019}. We consider two auxiliary problems with the exact same geometry as our original problem. We take these to be $(\rm i)$ two passive particles that trail one another with equal and constant velocities, and $(\rm ii)$ two passive particles that approach one another with equal speeds and opposite velocities. One can then find the velocities of each particle in terms of the auxiliary problems (see \cite{sharifi2016,papavassiliou2017,yang2019} for the derivation) as
\begin{align}
\label{final}
\bigg[\begin{matrix} 
\bm{V}_1 \\
\bm{V}_2 
\end{matrix}\bigg]=-\bigg[\begin{matrix} 
{\bm{F}_1^{(\rm i)}} & {\bm{F}_2^{(\rm i)}} \\
{\bm{F}_1^{(\rm ii)}} & {\bm{F}_2^{(\rm ii)}}
\end{matrix}\bigg]^{-1}\cdot\bigg[\begin{matrix} 
{\cal T}^{(\rm i)} \\
{\cal T}^{(\rm ii)}
\end{matrix}\bigg],
\end{align}
where ${\bm{F}_1^{(a)}}$ and ${\bm{F}_2^{(a)}}$ are the net hydrodynamic forces on the particles in the auxiliary problems,
${\cal T}^{(a)}=\sum_{k=1}^2\left\langle\bm{n}_k\cdot\boldsymbol{\sigma}^{(a)}\cdot\bm{v}_k^{\rm s}\right\rangle_{{\cal S}_k}$, $\langle \cdot \rangle$ denotes the surface integral, and $a\in\{{\rm i},{\rm ii}\}$. Noting that the details of the auxiliary problems are well-documented \cite{stimson1926,maude1961,spielman1970,happel1981}, the instantaneous velocities of the spheres can be simply found from Eq. \eqref{final}.

Due to the continuum framework adopted for the chemical interactions and the assumption of constant normal fluxes at the surface of each sphere, we cannot allow the gap size to reach zero. We thereby define a cut-off gap size of $\Delta/R=0.001$, below which we assume the particles are essentially in contact. Given that the exact solutions are expressed as sums of infinite series, we truncate them such that the convergence of the solution is ensured \cite{yariv2003}.

\begin{figure}
\begin{center}
\includegraphics[width=0.85\columnwidth]{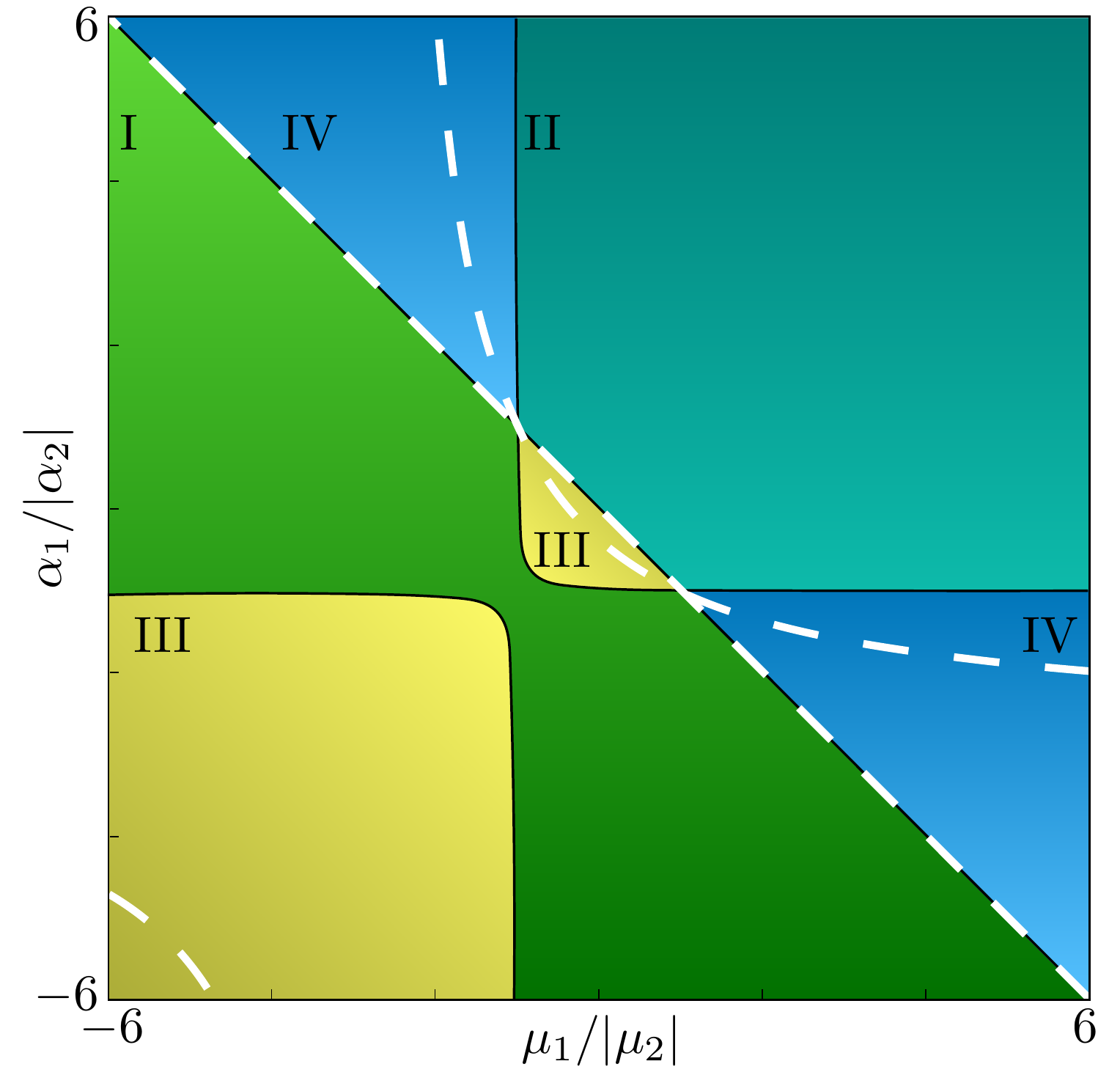}
\caption{The regime diagram for the interactions of two chemically-active particles in the activity-mobility parameter space, when ${\alpha}_2{\mu}_2>0$. If ${\alpha}_2{\mu}_2<0$, the map needs to be inverted, namely regime I changes to regime II, regime III changes to regime IV, and vice versa. The dashed lines show the regime boundaries when the hydrodynamic interactions are neglected.}
\label{phase}
\end{center}
\end{figure}

There are four possibilities for the relative motion of two chemically-active spheres: The two particles collapse onto one another and form a complex (regime I), move away and separate (regime II), reach a stable equilibrium (bound-state) at a certain gap size (regime III), or develop a critical gap size above which they move apart and below which they aggregate (regime IV). As shown in Fig.~\ref{regimes}, while in regime I and II the relative longitudinal velocity (i.e., $\left(\bm{V}_1-\bm{V}_2\right)\cdot\bm{e}$) is strictly negative or positive, in regimes III and IV, it becomes zero at a nonzero gap size. Since the phoretic interactions of two chemically-active particles are nonreciprocal, being in regimes I or II does not necessitate having mutual attraction or repulsion. Instead, we can have frustrated cases such as particle 1 being attracted to particle 2, while particle 2 is repelled by particle 1, etc \cite{soto2014,soto2015,canalejo2019}. As shown in the insets of Fig~\ref{regimes}, similarly in regimes III and IV, switching from mutual attraction to repulsion can happen due to the change of direction or magnitude of the interaction. In the example given for regime III in Fig~\ref{regimes}(c), the direction of both particles flip when the gap size exceeds the fixed-point value, while in the case depicted for regime IV in Fig~\ref{regimes}(d), the direction and strength change together to reverse the relative motion.

We recall that in the far-field approximation the sign of ${\alpha}_2 {\mu}_1+{\alpha}_1 {\mu}_2$ determines the behavior of the system (regime I, if negative, and regime II, if positive). To similarly find a criterion for these four regimes in the exact description of the problem, we need to further investigate the relative motion. Since the spheres are of equal radii, we have $\bm{F}_1^{(\rm i)}=\bm{F}_2^{(\rm i)}$ and $\bm{F}_1^{(\rm ii)}=-\bm{F}_2^{(\rm ii)}$. The relative velocity is then 
\begin{equation}
\label{relative}
\bm{V}_1-\bm{V}_2=\frac{\bm{e}}{|\bm{F}_1^{(\rm ii)}|}\left(\left<\sigma_{1}^{(\rm ii)}{v}_1^{\rm s}\right>_{{\cal S}_1}+\left<\sigma_{2}^{(\rm ii)}{v}_2^{\rm s} \right>_{{\cal S}_2}\right),
\end{equation}
where $\sigma_{k}^{(\rm ii)}=\bm{n}_k\cdot\boldsymbol{\sigma}^{(\rm ii)}\cdot\bm{t}_k$, $\bm{v}^{\rm s}_k=v^{\rm s}_k\bm{t}_k$, and $\bm{t}_k$ is a unit vector tangential to the surface of the $k$th sphere (respecting the axial symmetry) \cite{expl}.

Due to the linearity of the chemical field equations and the symmetric geometry, we can write $C(\bm{x})=\alpha_1\mathcal{G}_1(\bm{x}-\bm{x}_1)+\alpha_2\mathcal{G}_2(\bm{x}-\bm{x}_2)$, where $\bm{x}_k$ represents the center of sphere $k$, and $\mathcal{G}_2$=$\mathcal{G}_1^*$ with $\left(\cdot\right)^*$ denoting a reflection transformation with respect to the the symmetry plane perpendicular to $\bm{e}$. The slip velocities are then found as $\bm{v}_k^{\rm s}=\mu_k\alpha_k\bm{\nabla}^k_\parallel \mathcal{G}_k(R\bm{n}_k)+\mu_k\alpha_l\bm{\nabla}^k_\parallel \mathcal{G}_l(R\bm{n}_k+\bm{x}_k-\bm{x}_l)$, where $\bm{\nabla}_\parallel^k=\left(\bm{I}-\bm{n}_k\bm{n}_k\right)\cdot\bm{\nabla}$, and $k,l \in\{1,2\}$ in a mutually exclusive manner. The slip velocity of each particle comprises a self-generated contribution and an induced one from the neighboring particle. Since the particles considered here are chemically isotropic, the former contribution can only be nonzero due to the near-field effects, and thus vanishes once the particles are far from one another. Combining the above definitions with Eq. \eqref{relative}, and defining $\mathcal{N}=\left\langle \sigma_{k}^{(\rm ii)} \bm{\nabla}^k_\parallel \mathcal{G}_k(R\bm{n}_k)\cdot\bm{t}_k  \right\rangle_{{\cal S}_k}$, and $\mathcal{F}=\left\langle \sigma_{k}^{(\rm ii)} \bm{\nabla}^k_\parallel \mathcal{G}_l(R\bm{n}_k+\bm{x}_k-\bm{x}_l)\cdot\bm{t}_k  \right\rangle_{{\cal S}_k}$, we obtain
\begin{align}
\label{sign}
 \bm{V}_1-\bm{V}_2=\frac{\mathcal{F} \bm{e}}{|\bm{F}_1^{(\rm ii)}|}\left[\left(\alpha_1\mu_1 + \alpha_2\mu_2\right)\varepsilon+ \left(\alpha_1\mu_2 + \alpha_2\mu_1\right)\right],
 \end{align}
where $\varepsilon=\mathcal{N}/\mathcal{F}$ characterizes the importance of the near-field hydrodynamic and chemical interactions and is solely a function of the gap size. As we will show later, $\mathcal{N}$ and $\mathcal{F}$ vary monotonically with the gap size, and asymptotically approach zero as $\Delta\rightarrow\infty$. Consequently, they always maintain the same sign (i.e. never cross zero), which we choose to be positive by definition; $\mathcal{N},\mathcal{F}>0$. Thus, the sign of $\left(\alpha_1\mu_1 + \alpha_2\mu_2\right)\varepsilon+ \left(\alpha_1\mu_2 + \alpha_2\mu_1\right)$ determines the attractive or repulsive nature of the overall phoretic interaction, and it may change depending on the gap size, giving rise to the appearance of regimes III and IV. This expression suggests that the qualitative role of the near-field effects---which may (or may not) oppose the far-field attraction or repulsion---is simply due to a self-generated effect, which is the gradient in the chemical field induced by the particle itself, due to the presence of its neighbor. The neighboring particle can then be simply interpreted as a passive boundary which introduces a geometrical asymmetry in the otherwise isotropic chemical field. We note that this is similar to how a substrate screens the motion of an adjacent phoretic particle, for which the existence of a fixed-point has also been observed \cite{Uspal2015,mozaffari2016}. As we show in the Supplemental Material \cite{supp_2body}, one can find a similar interplay between far-field and near-field effects for a Janus particle near a substrate. The far-field effect in this case is replaced by the self-propulsion, and the near-field role is played by the substrate-induced motion.

The chemical gradient terms embedded in $\mathcal{N}$ and $\mathcal{F}$ are essentially the slip velocities of sphere $k$ when $\alpha_l=0$, and $\alpha_k=0$, respectively. Thus, in both cases, their magnitude can only decay by increasing the gap size. Given that in the approaching auxiliary problem, $\sigma_k^\text{(ii)}$ also decays with $\Delta$ \cite{happel1981}, we expect both $\mathcal{N}$ and $\mathcal{F}$ to decay monotonically with respect to $\Delta$. Interestingly, we find that the variation of their ratio, $\varepsilon$, with $\Delta$ is also monotonic. Therefore, to determine the regime of the system, it suffices to compare the sign of Eq. \eqref{sign} when the particles are very close ($\Delta\rightarrow0$), and when they are very far ($\Delta\rightarrow\infty$). For the latter $\varepsilon\rightarrow 0$, which recovers the far-field approximation criterion for attraction and repulsion. When the particles are close, however, we find $\varepsilon\approx 0.97$. Using these two bounding cases, we can construct the regime diagram in the activity-mobility parameter space, as shown in Fig.~\ref{phase}. It is worth noting that for chemically-identical particles (${\alpha}_1={\alpha}_2$ and ${\mu}_1={\mu}_2$), since $1+\varepsilon>0$, the system is in regime I when ${\alpha}_k {\mu}_k<0$, and regime II when ${\alpha}_k {\mu}_k>0$, which is in agreement with the prediction of the far-field approach.

\begin{figure}
\begin{center}
\includegraphics[width=\columnwidth]{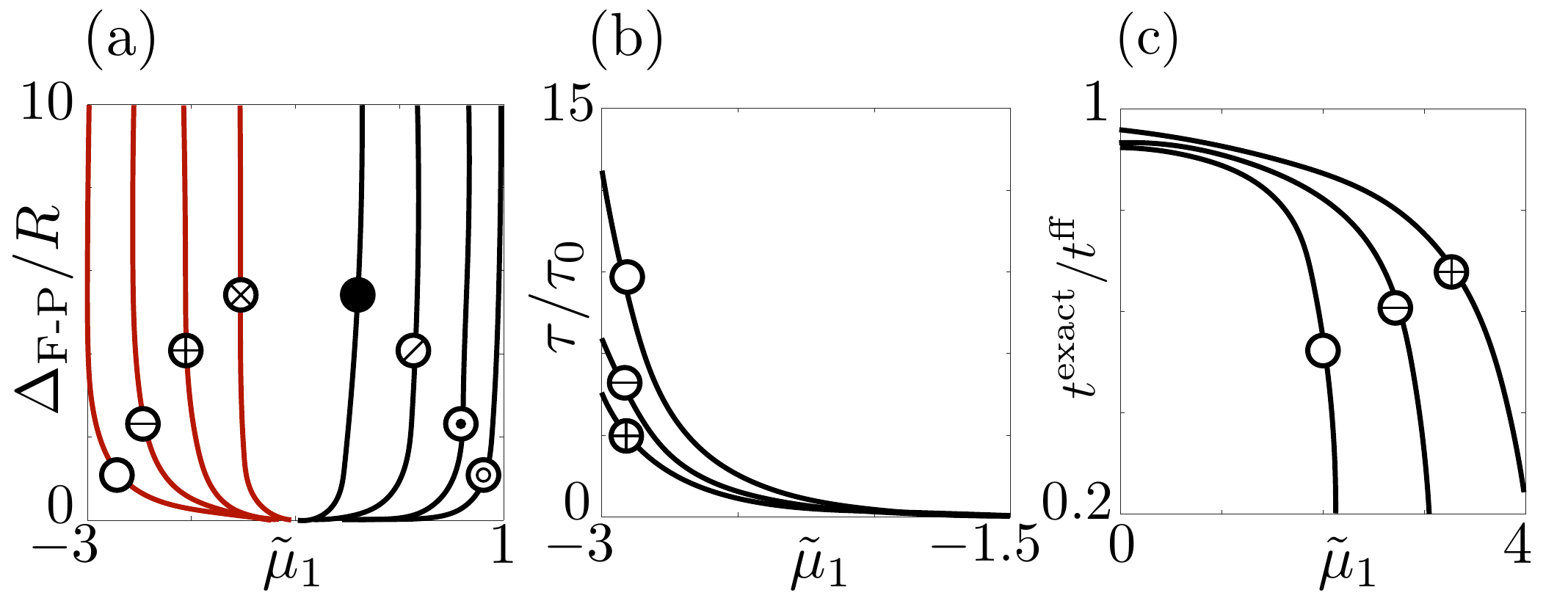}
\caption{(a) The variation of fixed-point location ($\Delta_\text{F-P}$) as a function of $\tilde{\mu}_1$ for $\tilde{\mu}_2=1$, and $\tilde{\alpha}_2=1$. The lines in red are unstable fixed-points with $\tilde{\alpha}_1=3$ ($\circ$), $\tilde{\alpha}_1=2.5$ ($\ominus$), $\tilde{\alpha}_1=2$ ($\oplus$) and $\tilde{\alpha}_1=1.5$ ($\otimes$). Those in black are stable fixed-points with $\tilde{\alpha}_1=0.5$ ($\bullet$), $\tilde{\alpha}_1=0$ ($\oslash$), $\tilde{\alpha}_1=-0.5$ ($\odot$) and $\tilde{\alpha}_1=-0.75$ ($\circledcirc$). (b) The mean first-passage time for particles in regime IV, in units of  $\tau_0=R^2/D_c^\infty$, corresponding to ${D}_c^\infty/RV_0=0.05$, where $D_c^\infty=D_c\left(\Delta\rightarrow\infty\right)$. Here, $\tilde{\alpha}_2=1$, and $\tilde{\mu}_2=1$, with $\tilde{\alpha}_1=3.5$ ($\circ$), $\tilde{\alpha}_1=4$ ($\ominus$), and $\tilde{\alpha}_1=4.5$ ($\oplus$). (c) The ratio of the collision time obtained from the exact solution, $t^\text{exact}$, and the corresponding value from the far-field solution $t^\text{ff}$ for $\tilde{\alpha}_2=1$ and $\tilde{\mu}_2=1$, with $\tilde{\alpha}_1=-2.1$ ($\circ$), $\tilde{\alpha}_1=-3.1$ ($\ominus$), and $\tilde{\alpha}_1=-4.1$ ($\oplus$).}
\label{fig3}
\end{center}
\end{figure}

In the absence of hydrodynamic interactions, the translational velocity of each particle is governed by $\bm{V}_k=-\frac{1}{4\pi R^2}\left<\bm{v}_k^{\rm s}\right>_{\mathcal{S}_k}$ \cite{anderson1989}. By repeating the same steps, we find the following expression for the relative velocity
\begin{align}
\label{sign2}
 \bm{V}_1-\bm{V}_2=\frac{\mathcal{F}_{\text{ch}} \bm{e}}{4\pi R^2}\left[\left(\alpha_1\mu_1 + \alpha_2\mu_2\right)\varepsilon_{\text{ch}}+ \left(\alpha_1\mu_2 + \alpha_2\mu_1\right)\right],
 \end{align}
where $\varepsilon_{\text{ch}}=\mathcal{N}_{\text{ch}}/\mathcal{F}_{\text{ch}}$ now represents the relative importance of the near-field chemical interactions, with $\mathcal{N}_{\text{ch}}=(-1)^k \langle \bm{\nabla}^k_\parallel \mathcal{G}_k(R\bm{n}_k)  \cdot\bm{t}_k \rangle_{{\cal S}_k}$, and $\mathcal{F}_{\text{ch}}=(-1)^{k+1}\langle \bm{\nabla}^k_\parallel \mathcal{G}_l(R\bm{n}_k+\bm{x}_k-\bm{x}_l)\cdot\bm{t}_k \rangle_{{\cal S}_k}$. When $\Delta\rightarrow \infty$ we find $\varepsilon_{\text{ch}}=0$, and $\Delta\rightarrow 0$ yields $\varepsilon_{\text{ch}}\approx  0.37$, indicating that the chemical interactions alone can capture all the regimes, but the boundaries differ from those of the complete solution, as shown by the dashed lines in Fig.~\ref{phase}. We also report that the emergence of a fixed-point cannot be captured when both chemical and hydrodynamic interactions are probed using only far-field approximations, emphasizing that the near-field effects are necessary for capturing regimes III and IV.

Note that regimes III and IV are located between regimes I and II, suggesting that the transition from the fully attractive mode to the fully repulsive mode is not abrupt. This behavior can be better explained through the variations of the fixed-point location ($\Delta_\text{F-P}$) in these regimes, as shown in Fig.~\ref{fig3}(a). In regime III, in which the system exhibits attractive behavior only when $\Delta>\Delta_\text{F-P}$, moving towards regime II results in an increase in $\Delta_\text{F-P}$, essentially expanding the range of the repulsive region. Upon reaching regime II, $\Delta_\text{F-P}\rightarrow\infty$, indicating that the system no longer has a fixed-point (and, consequently, an attractive region). On the other end, moving towards regime I shrinks the repulsive region by decreasing $\Delta_\text{F-P}$ to zero. Similarly, for the unstable fixed-point in regime IV, $\Delta_\text{F-P}\rightarrow0$ when approaching regime II and $\Delta_\text{F-P}\rightarrow\infty$ when reaching regime I. 

The complexes formed in regime IV are metastable, and they can dissociate in the presence of noise (e.g., thermal fluctuations). We investigate this activation process by evaluating the mean first-passage time, $\tau$, which characterizes the time required for the particles to escape a potential barrier.
To this end, we can write a Fokker-Planck equation for the relative distance with the drift velocity given by 
$-\frac{\text{d} \mathcal{U}}{\text{d} \Delta} \bm{e}=\bm{V}_1-\bm{V}_2$, with an effective diffusion coefficient $D_c(\Delta)$ that measures the strength of noise. Using the standard solution for such a process, we find
\begin{align}
\tau=\int_0^{\Delta_{\text{F-P}}}  \frac{\text{d}r_1}{r_1^{2}} \,e^{\psi(r_1)}\int_0^{r_1}\frac{ r_2^2 \text{d}r_2}{D_c(r_2)}  \,e^{-\psi(r_2)},
\end{align}
where $\psi(r)=\int_0^r \frac{\text{d}r’}{D_c(r’)}\frac{\text{d}\mathcal{U}}{\text{d}r’}$.
The result is shown in Fig.~\ref{fig3}(b) for different parameter values. We observe that $\tau$ increases rapidly when the system is closer to regime II than regime I, indicating that the systems with larger $\Delta_\text{F-P}$ (or attractive region) are more stable. 

Finally, we investigate how the near-field effects modify the dynamics of the particles when the behavior of the system is correctly predicted by the far-field limit, namely in regimes I and II. We probe this difference by comparing the collision time of the particles in regime I, as obtained from the exact solution ($t^\text{exact}$), and the far-field approximation ($t^\text{ff}$); see Fig.~\ref{fig3}(c). We observe that the timescales can differ significantly, particularly close to the transition boundaries. As Fig.~\ref{fig3}(c) shows, the far-field solution can substantially overestimate the collision time.

In conclusion, we have shown that a system of two chemically-active particles can exhibit four different types of behavior, of which only two can be qualitatively captured using the far-field approximation.
Our results highlight that accounting for near-field effects can be crucial when analyzing phoretic systems. We have provided a simple expression for determining the regime of a system, which can be used as a benchmark for the far-field approximations used in systems with pair-interactions. For instance, regime III may be treated as regime I, but with effective radii for the particles that correlate with the equilibrium gap size. Due to the generality of our framework, one can extend this approach and study anisotropic particles \cite{michelin2017}, include steric effects \cite{varma2018}, introduce a nearby boundary \cite{Uspal2015,mozaffari2016}, and resolve non-axisymmetric interactions \cite{sharifi2016,saha2019}. Given that the near-field corrections in many-body solvers are often based on pair-interactions \cite{brady1988}, our result can be useful for introducing the near-field phoretic interactions in far-field-based approaches \cite{varma2019}.


\begin{thebibliography}{64}
\bibitem [{exp()}]{expl}%
  \BibitemOpen
  \href@noop {} {}\bibinfo {note} {Note that only the approaching auxiliary
  problem is needed to determine the relative motion of the spheres. One can
  similarly show that the center-of-mass motion is governed by the trailing
  problem.}\BibitemShut {Stop}%
\end{thebibliography}
\end{document}


\title{Supplemental Material -- Exact phoretic interaction of two chemically-active particles}
\author{Babak Nasouri}
\affiliation{Max Planck Institute for Dynamics and Self-Organization (MPIDS), 37077 Goettingen, Germany}
\author{Ramin Golestanian}
\email{ramin.golestanian@ds.mpg.de}
\affiliation{Max Planck Institute for Dynamics and Self-Organization (MPIDS), 37077 Goettingen, Germany}
\affiliation{Rudolf Peierls Centre for Theoretical Physics, University of Oxford, Oxford OX1 3PU, United Kingdom}
\date{\today}

\maketitle
\section{chemically-active particle near a flat substrate}
Here we study the behavior of a chemically-active particle near a flat substrate, and show that similar to the system of two particles discussed in the main article, here as well the system can have a fixed-point due to near-field effects. To this end, we consider a single sphere of radius R whose center is at distance $(\Delta+R)$ from a flat substrate. The system is axisymmetric along the axis $\bm{e}$, where $\bm{e}$ is a unit vector perpendicular to the plane of the substrate, pointing towards the center of the sphere. In the absence of advection (vanishingly small P\`eclet number), the concentration field is governed by $\nabla^2C=0$, subject to $-D\bm{n}\cdot\bm{\nabla}C=\alpha$ at the surface of the sphere and $\bm{e}\cdot\bm{\nabla}C=0$ on the substrate, where $\bm{n}$ is a unit vector normal to the surface of the sphere, and $D$ is the diffusion coefficient for the chemical. Here, $\alpha$ is the surface activity which may vary along the surface of the sphere, and is axisymmetric with respect to $\bm{e}$. The perturbation in the concentration field can give rise to a local slip velocity on the sphere, which is governed by $\bm{v}^\text{s}=\mu\left(\bm{I-nn}\right)\bm\cdot\bm{\nabla}C$, where $\mu$ is the mobility, and we assume is constant across the surface of the sphere.

\begin{figure}
\begin{center}
\includegraphics[width=0.6\columnwidth]{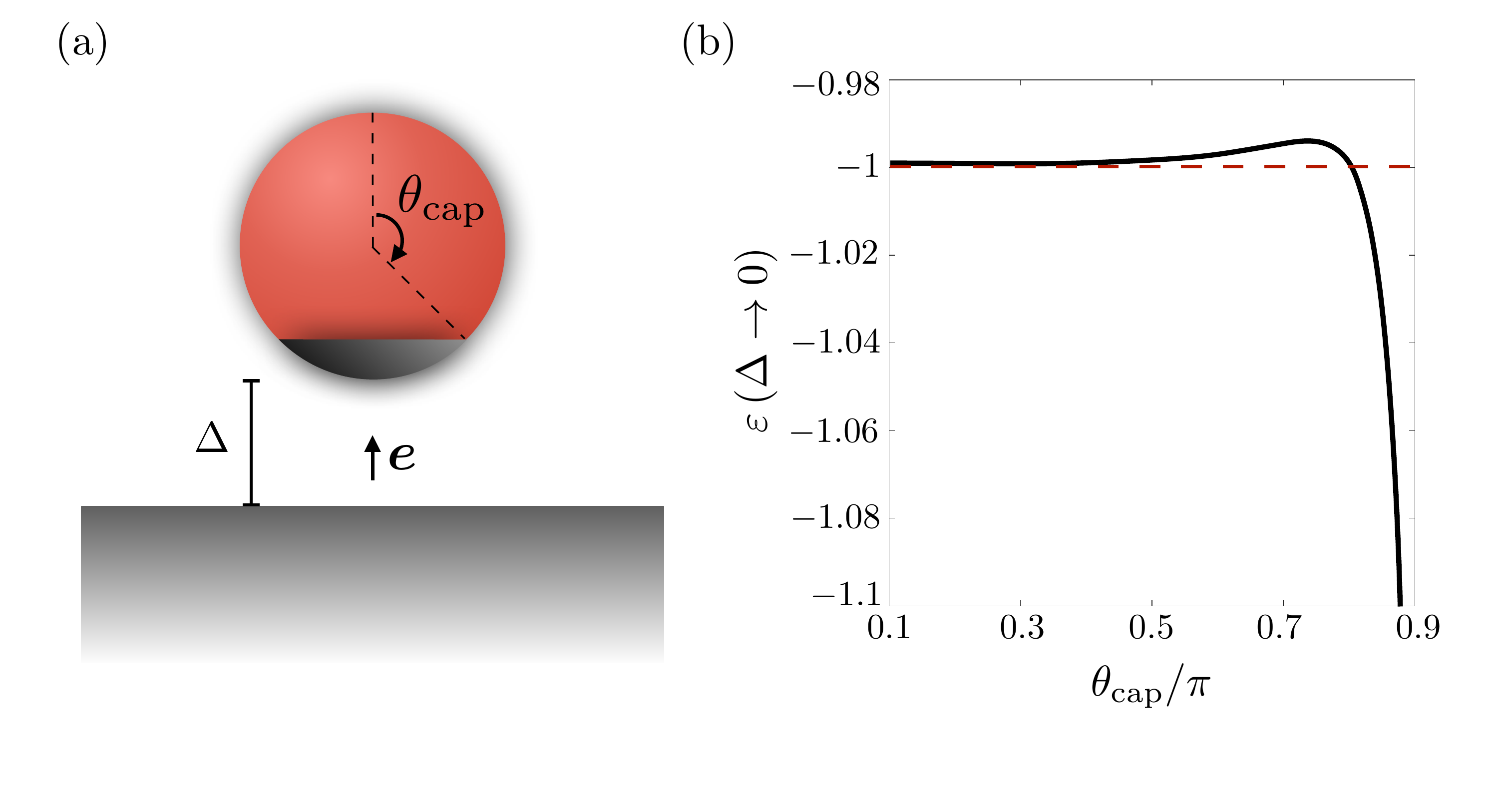}
\caption{(a) Schematic of the considered Janus particle. The bottom compartment (grey) is inert, and the top compartment (red) is chemically-active. (b) Variation of $\varepsilon\left(\Delta\rightarrow0\right)$ with the coating-coverage angle $\theta_\text{cap}$. The dashed-line shows the value $-1$.}
\label{fig}
\end{center}
\end{figure}

Due to the linearity of the chemical field equations, we can decompose the concentration field as
\begin{align}
C(\bm{x})=C_1(\bm{x}) + C_2(\bm{x}) \tag{S1},
\end{align}
where $C_1$ indicates the concentration field induced purely by the sphere in the absence of the substrate, while $C_2$ accounts for the effect of the substrate. This way, the slip velocity can be found as
\begin{align}
\label{slip}
\bm{v}^\text{s}=\mu\left(\bm{\nabla}_{\parallel}C_1 +\bm{\nabla}_{\parallel}C_2 \right)\tag{S2},
\end{align}
where $\bm{\nabla}_{\parallel}=\left(\bm{I-nn}\right)\cdot\bm{\nabla}$. To resolve the hydrodynamic interactions, we once again use the reciprocal theorem, this time choosing the translational motion of a passive sphere towards a flat wall as the auxiliary problem \cite{papavassiliou2015,mozaffari2016}. The velocity of the sphere is then found
\begin{align}
\bm{V}=\frac{-\left<\bm{n}\cdot\hat{\bm{\sigma}}\cdot\bm{v}^\text{s}\right>\bm{e}}{\hat{\bm{F}}\cdot\bm{e}}\tag{S3},
\end{align} 
where $\hat{\bm{F}}$ is the net hydrodynamic force on the sphere in the auxiliary problem, and $\hat{\bm{\sigma}}$ is its stress field, both of which can be found explicitly from the work of \citet{brenner1961}. Here, $\left<\cdot\right>$ denotes integration over the surface of the sphere. Using Eq. \eqref{slip}, we can rewrite the translational velocity as
\begin{align}
\label{generic}
\bm{V}=\frac{-\mu\bm{e}}{\hat{\bm{F}}\cdot\bm{e}}\left(\mathcal{F}+\mathcal{N}\right)\tag{S4},
\end{align} 
where $\mathcal{F}=\left<\bm{n}\cdot\hat{\bm{\sigma}}\cdot\bm{\nabla}_{\parallel}C_1\right>$, and $\mathcal{N}=\left<\bm{n}\cdot\hat{\bm{\sigma}}\cdot\bm{\nabla}_{\parallel}C_2\right>$. Note that here, $\mathcal{F}$ represents the translational motion induced by the particle in the absence of the substrate (i.e., self-propulsion), and $\mathcal{N}$ denotes the correction due to the presence of the substrate. When the gap size is large $\Delta\rightarrow\infty$, the effect of the substrate vanishes ($\mathcal{N} \rightarrow 0$), and the motion of the particle is dominated by the self-induced thrust. But when the gap size is small, both $\mathcal{N}$ and $\mathcal{F}$ contribute to the motion, and the propulsion direction is decided by their interplay. 

We evaluate  Eq. \eqref{generic} for two types of particles: isotropically coated, and partially coated (i.e., Janus). For an isotropically-coated particle, since the concentration gradient in the absence of the substrate is completely isotropic, we find $\mathcal{F}=0$, indicating that the motion is purely due to the presence of a substrate. For such a system, $|\mathcal{N}|$ monotonically decreases to zero as the gap size increases, and so no fixed-point can be seen.

We now consider a Janus particle which has two compartments, inert and active, as shown in Fig.~\ref{fig}(a). We assume the inert compartment faces the substrate, and the active one has a homogeneous activity $\alpha$. We define $\theta_\text{cap}$ as a measure of the coating coverage, which varies between $0$ (completely inert) and $\pi$ (isotropically coated). For $\theta_\text{cap}\neq\{0,\pi\}$, the particle can self-propel and so $\mathcal{F}\neq0$. We can then rewrite Eq. \eqref{generic} as
\begin{align}
\bm{V}=\frac{-\mu\mathcal{F}\bm{e}}{\hat{\bm{F}}\cdot\bm{e}}\left(1+\varepsilon\right)\tag{S5},
\end{align} 
where $\varepsilon=\mathcal{N}/\mathcal{F}$ characterizes the role of near-field effects and is solely function of the gap size. When the gap size is large, we expectedly find $\varepsilon\rightarrow 0$. We may then evaluate $\varepsilon$ when $\Delta\rightarrow0$ for different values of $\theta_\text{cap}$; see Fig~\ref{fig}(b). We note that, for $\theta_\text{cap}\gtrsim0.8\pi$, the value of $\varepsilon$ falls below $-1$, indicating that the system has a fixed-point. Depending on the sign of $\alpha\mu$, this fixed-point can be stable ($\alpha\mu>0$), or unstable ($\alpha\mu<0$). For instance, for the specific Janus particle considered in Ref. \cite{Uspal2015} ($\theta_\text{cap}\approx0.856\pi$, $\alpha\mu>0$), we find $\varepsilon\approx-1.04$ when $\Delta\rightarrow0$, meaning that the system has a stable fixed-point, which is in agreement with the computational results in that reference.

Recall that here there are two competing contributions to the net motion: one accounts for the self-propulsion and the other captures the role of the substrate. For the system of two isotropic particles discussed in the main article, there is no self-propulsion, nor is there a substrate. Instead, the self-propulsion contribution is replaced by the motion originated from the concentration gradient of the neighboring particle, and the role of substrate is substituted by a self-generated term which is nonzero due to the passive presence of the neighboring particle.